\documentclass[useAMS,usenatbib]{mn2e}

\usepackage{graphicx} \usepackage{epsf} \usepackage{mathrsfs}
\title[Temporal variations in the acoustic signal from faculae]{Temporal variations in the acoustic signal from faculae}
\author[C. Karoff]{C. Karoff\thanks{E-mail: karoff@phys.au.dk}\\
Department of Physics and Astronomy, Aarhus University, DK-8000 Aarhus C, Denmark\\}
\begin{document}

\date{Accepted ... Received ...}

\pagerange{\pageref{firstpage}--\pageref{lastpage}} \pubyear{2012}

\maketitle

\label{firstpage}

\begin{abstract}
The integrated brightness of the Sun shows variability on time-scales from minutes to decades. This variability is mainly caused by pressure mode oscillations, by granulation and by dark spots and bright faculae on the surface of the Sun. By analyzing the frequency spectrum of the integrated brightness we can obtain greater knowledge about these phenomena. It is shown how the frequency spectrum of the integrated brightness of the Sun in the frequency range from 100 to 3200 $\mu$Hz shows clear signs of both granulation, faculae and p-mode oscillations and that the measured characteristic time-scales and amplitudes of the acoustic signals from granulation and faculae are consistent with high-resolution observations of the solar surface. Using 13 years of observations of the Sun's integrated brightness from the VIRGO instrument on the $SOHO$ satellite it is shown that the significance of the facular component varies with time and that it has a significance above 0.99 around half the time. Furthermore, an analysis of the temporal variability in the measured amplitudes of both the granulation, faculae and p-mode oscillation components in the frequency spectrum reveals that the amplitude of the p-mode oscillation component shows variability that follows the solar cycles, while the amplitudes of the granulation and facular components show signs of quasi-annual and quasi-biennial variability, respectively.
\end{abstract}


\begin{figure}
\includegraphics[width=\columnwidth]{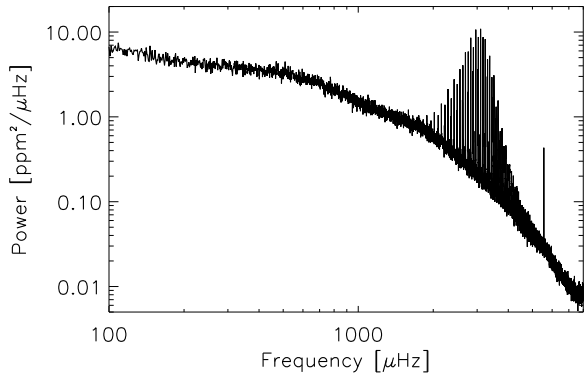}
\caption{The solar acoustic spectrum. The spectrum was calculated as the mean of 119 one months spectra (see text). Both the granulation component (around 800 $\mu$Hz), the facular component (around 1800 $\mu$Hz) and the p-mode oscillations component (around 3000 $\mu$Hz) are seen as changes in the slope of the spectrum. The different components are more clearly marked in Fig.~2. It is also seen that the spectrum declines faster at frequencies higher than the p-mode oscillations or the atmospheric acoustic cut-off frequency. The spectrum was calculated from 13 years of observations of the Sun's integrated brightness from the VIRGO instrument on the $SOHO$ satellite and the peak at 5555 $\mu$Hz is an artefact.}
\end{figure}
\section{Introduction}
The frequency spectrum of observations of the integrated brightness of the Sun is often referred to as the solar acoustic spectrum \citep{1976ApJ...209L..39I}, as it is mainly used to study pressure or sound waves inside the Sun. These sound waves can be observed on the surface of the Sun as oscillation pattens which can be represented as spherical harmonics of the eigenfunctions of the sound waves inside the Sun. These sound waves lead to a series of distinct peaks in the solar acoustic spectrum (see Fig.~1) and by studying the frequencies, amplitudes and line-widths of these peaks we can through helioseismology obtain knowledge of the physics that takes place inside the Sun \citep[see e.g.][]{1996Sci...272.1296G}, but more information are hidden in the solar acoustic spectrum. At lower frequency the solar acoustic spectrum contains signals from spots, granulation and faculae. In fact the acoustic signal from granulation and faculae continues up to the frequencies of the p-mode oscillations and even to higher frequencies and therefore this signal is often referred to as the acoustic background. Studies of Sun-like stars have shown that other stars show a similar acoustic background \citep[see e.g.][]{2008Sci...322..558M, 2010ApJ...713L.169C}. 

\citet{1985shpp.rept..199H} could be the first to suggest that the solar acoustic background was modelled with a number of exponentially decaying functions. The idea is that the signal from e.g. granulation is understood as a random signal with some memory and thus its autocorrelation function should be given by an exponentially decaying function defined by a characteristic time-scale and an amplitude \citep{2007arXiv0710.3378B}. Long uninterrupted observations from the South Pole and from the $SOHO$ satellite have later revealed that the solar acoustic background contains signals from many more phenomena than just granulation. \citet{1993ASPC...42..111H} identified a total of 6 difference components in the acoustic spectrum from observations of the Ca~{\sc ii}~K line in the solar atmosphere from the South Pole. The identified components were, from lower to higher frequency: non-periodic fluctuations attributed to active region evolution; non-periodic fluctuations attributed to granulation overshoot; non-periodic fluctuations attributed to chromospheric bright points; periodic fluctuations attributed to the photosphere (i.e. p-mode oscillations) and periodic fluctuations attributed to the chromosphere (i.e. chromospheric oscillations). Whereas the signals from granulation and p-mode oscillations are reported in all studies, the signals associated with the chromosphere are only reported in the study by \citet{1993ASPC...42..111H}, as this is the only study that analyzes the acoustic background in observations of the Ca~{\sc ii} lines. The feature between the granulation signal and the p-mode oscillations has on the other hand also been reported in most studies, especially studies based on intensity measurements, but an agreement on the origin of this features has not been reached. 

\citet{2004ESASP.538..215A} analyzed observations of solar irradiance from the VIRGO instrument on $SOHO$ and identified not only a granulation component in the frequency spectrum, but also components from active regions, super- and meso-granulations and bright points. The characteristic time-scale of the component that \citet{2004ESASP.538..215A} identified as bright points is similar to the characteristic time-scale of the component that \citet{1993ASPC...42..111H} identified as chromospheric bright points. \citet{2005A&A...443L..11V} identified the same component in integrated light observations from VIRGO, but speculated that this component might be due to a second granulation population -- as shall be showen later this is not likely.

Bright points and faculae are related phenomena and they are both related to the magnetic network on the solar surface. The difference is that whereas bright points are mainly seen in the inter-granular lanes, faculae are seen inside the granules and that whereas bright points can maintain their brightness over several tens of minutes \citep{2007ApJ...661.1272B}, faculae change significantly on granular evolution time-scales \citep{2006ApJ...646.1405D}. For this reason it is more likely that the component between the granulation and p-mode oscillation components is caused by faculae rather than bright points.

In the picture of \citet{2007ApJ...661.1272B} faculae are deffined as "{\it the edges of granules} seen through the "forest" of the magnetic field in plages and the network". The faculae does in other words occur in a localized region of the granules and it is thus clear that the characteristic time-scale of faculae will be a bit shorter than that of granulation. There are no quantitative studies that directly relates the time-scale of granulation to the time-scale of faculae based on high-resolution observations. A qualitative argument for that the characteristic time-scale of faculae is shorter than that of granulation is given in the small movie in Fig.~9 of \citet{2006ApJ...646.1405D}. This small movie of a 3D radiative magneto convective simulation shows 3 distinct granules that live throughout the full length of the movie (330 sec) and a number of faculae on these granules. If one look e.g. at the facular at (0$\farcs$3,1$\farcs$4) it is clearly visible at $t = 0$ sec, but completely gone at $t = 90$ sec. At $t = 120$ sec a new facular starts to evolve just to the right of where the other was located and at $t = 210$ sec, this facular has its peak brightness. In other words, though the evolution of faculae follows the evolution of the granules it is clear from the movie that the same granule can, during its lifetime, contain a number of faculae -- i.e. the characteristic time-scale of faculae is shorter than that of granulation.

A few studies \citep{1998mons.proc...59T, 2006A&A...445..661L, 2009A&A...506..167L} have used hydrodynamical models of stellar atmospheres to simulate acoustic spectra of Sun-like stars, but so far no such studies have been made using {\it magneto}hydrodynamical models, which is needed in order to reproduce the facular component.

In this paper we will use observations from the green channel on the VIRGO instrument on $SOHO$ \citep{1997SoPh..170....1F} to analyze the acoustic signature of faculae. We will show how the characteristic time-scale and amplitude that we measure of this component in the acoustic background are consistent with measurements based on high-resolution observations of the solar surface. We also analyze any temporal variations in the different components in the solar acoustic background -- including the facular component.

The paper is arranged as follows. In Section 2 we give a short theoretical description of the acoustic background and present the model that we use to model the granulation, the faculae and the p-mode oscillations components in the acoustic spectrum. This model is used in the analysis in Section 3, where we also describe a statistical test of the significant of the facular component in the acoustic background. Section 4 presents the results of the analysis and concluding remarks are found in section 5. The paper also includes an appendix that describe a test based on Monte Carlo simulations of the statistical method developed in Section 3.

\section{Theory}
The original formulation of the solar acoustic background by \citet{1985shpp.rept..199H} is the following:
\begin{equation}
f(\nu)=\frac{4\sigma^2\tau}{1+(2\pi\nu\tau)^2},
\end{equation}
where $f(\nu)$ is the power density at frequency $\nu$, $\sigma$ is the amplitude of the signal and $\tau$ is the characteristic time-scale. Eq. 1 is normalized to Parseval's theorem so that the variance of the signal in the time domain equals the sum of $f(\nu)$ over all frequencies:
\begin{equation}
\sigma^2=\int{f(\nu) \ d\nu},
\end{equation}
which means that $\sigma$ in eq. 1 does in fact reflects the variance of the signal in the time domain \citep{2007arXiv0710.3378B}. Note that every component in the acoustic background will contribute with the signal given by eq. 1 to the spectrum. The spectrum will thus be a sum of individual signals from i.e. granulation and faculae each defined by unique values of $\sigma$ and $\tau$.

\begin{figure}
\includegraphics[width=\columnwidth]{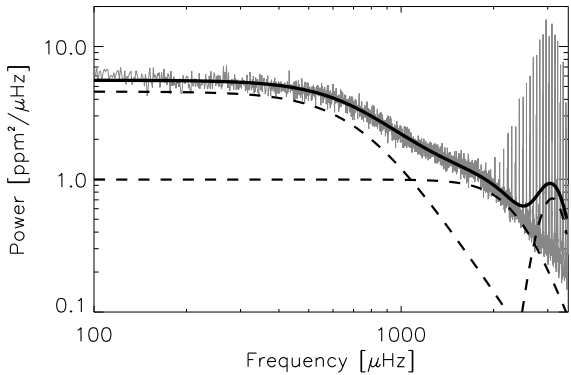}
\caption{The same a Fig.~1, but here with the model containing a facular component overlaid. The solid line shows the full model including all three components, whereas the dashed lines show the different components in the model (i.e. granulation, faculae and p-mode oscillations). The parameter values used in the model are given in Table 1.}
\end{figure}
Unfortunately, the original model by \citet{1985shpp.rept..199H} fails to reproduce the observed solar acoustic background for frequencies higher than the atmospheric acoustic cut-off frequency. The reason for this is that granulation cannot be modelled with turbulent cascades \citep{1997A&A...328..229N} as it is done in the drift model by \citet{1985shpp.rept..199H}. Turbulence shows a distribution with a slope of around $-2$ in power, convection on the other hand has a lower limit in the time domain on which changes can take place. This means that on small time-scales (or at high frequency) convection is not noisy whereas turbulence is and therefore the acoustic background decays with a slope smaller than $-2$. In the Sun the decay rate turns out to be around $-4$ at frequencies higher than the atmospheric acoustic cut-off frequency. Taking this into account the Harvey model extends to:
\begin{equation}
f(\nu)=\frac{4\sigma^2\tau}{1+(2\pi\nu\tau)^2+(2\pi\nu\tau)^4}.
\end{equation}
\citet{karoff2008} used this model to successful describe the granulation and facular components in the solar acoustic background between 400 $\mu$Hz and up to 8000 $\mu$Hz, which is close to the Nyquist frequency. 

\citet{1993ASPC...42..111H} used another approach to solve the problem with the fast decline at high frequency and considered the slope as a free parameter. When this is done the decline will often be faster then $-2$ and this will solve the problems at high frequency. In this way \citet{1993ASPC...42..111H} found slopes of $-5.6$ and $-5.0$ for the granulation and the bright point components, respectively. The slope gives the decay rate, which 'calibrates' the amount of memory in the process responsible for the noise \citep{1993ASPC...42..111H} and there is no reason that this value should be exactly $-2$, or that is should be the same for the different components. On the other hand, there is a physical argument why the slope should not be the same at low and high frequency (which were given above), so allowing the slope to be a free parameter only partly solves the problem.

\begin{figure}
\includegraphics[width=\columnwidth]{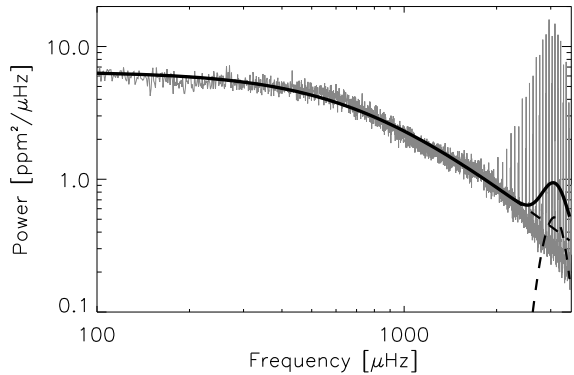}
\caption{The same as Fig.~2, but here the model does not contain a facular component. The parameter values used in the model are given in Table 1.}
\end{figure}
We have therefore chosen another approach and modelled only the lower frequency part of the spectrum (up to 3200 $\mu$Hz). This means that no extra component is needed in order to model the high-frequency part. The following model is thus used  to model the granulation and facular signals in the solar acoustic background between 100 and 3200 $\mu$Hz:
 \begin{equation}
 f(\nu)=\frac{4\sigma^2\tau}{1+(2\pi\nu\tau)^{\alpha}}.
 \end{equation}

The envelope of the p-mode oscillations shows variability on a weekly time-scale with an amplitude of 6.2\%, which is properly mainly due to changes in the damping rates of the oscillation modes \citep{2008ApJ...682.1370K}. The envelope of the p-mode oscillations therefore needs to be modelled along with the acoustic background in order to disentangle changes in the damping rates of the oscillation modes from changes in the acoustic background. A Gaussian envelope is therefore also included in the model in order to model the envelope of the p-mode oscillations, but as the high frequency part of the spectrum is not analyzed it is not necessary to include a white noise background. Of course the background will have some effect on the amplitudes that are measured of the granulation and facular components, but as the white noise level in the VIRGO observations is generally around three orders of magnitudes smaller than the granulation and facular amplitudes this effect is negligible. 

A Gaussian has here been chosen for the envelope of the p-mode oscillations, but one or two Lorentzians could also have been chosen \citep{1993ASPC...42..111H, 2008A&A...490.1143L}. This would how ever not affect the final results significantly and it is not clear which functional form that agrees best with the behavior of the Sun and other Sun-like stars in general \citep[see e.g. the simulations by][]{1999A&A...351..582H}.

Figs.~2~\&~3 show examples of fits to the solar acoustic background using the models described above either containing or not containing a facular component. The observations and the calculation of the spectrum are described in Section~4. 
 
\section{Analysis}
The data analysis consists of three steps; calculating the power density spectra, modelling the power density spectra and evaluating the significance of the different measured parameters and the goodness of the different models.

\subsection{Observations}
For the analysis we use a 13-year time series from the green channel on the VIRGO instrument of $SOHO$ \citep{2009A&A...501L..27F}. There are some gaps in this time series, especially around the $SOHO$ 'vacation' and recent measurements do suffer from increased noise level. In the analysis performed here we have divided the full 13-year time series into substrings of one month each. In order to get the lowest probability for artefacts such as gaps and increased noise levels to affect the data analysis we have performed a discrimination of both bad data points and bad substrings. Bad data points were identified by calculating a logarithmic running box variance log $\left(\sigma_i\right)$: 
\begin{equation}
{\rm log} \left( \sigma_i \right)= {\rm log}\left(\sqrt{\frac{1}{N}\sum_{j=i-w}^{j=i+w}\left[x_j-\mu \right]^2}\right),
\end{equation} 
where $x_j$ is the $j$th observation, $\mu$ is the mean value of the observations between $i-w$ and $i+w$ and $w$ is the width of the running box which was set equal to 10 points ($N$ = 21). Using this formulation all data points with a logarithmic running box variance that deviated more than 5$\sigma$ from the mean were then removed. This removed around 5\% of the data points. The reason for using the logarithmic variance is that many data points show artificially low scatter. 

After removing the bad data points all monthly substrings with a duty cycle lower than 95\% were also removed. This left us with 119 monthly substrings out of 147 possible. We did test that the relative high demands for selecting good data points did not affect the results. The test showed slightly divagating results when using either the 119 good monthly substrings or all the 147 substrings, but the differences were insignificant.

The power density spectra were calculated using least-squares \citep{1976Ap&SS..39..447L, karoff2008}. Each spectrum was normalized by the effective observation length given as the reciprocal of the area under the window function in order to convert the spectra into power density spectra. 

\subsection{Modelling the acoustic background}
Acoustic backgrounds have earlier been modelled by smoothing the spectra with a Gaussian running mean, with a width of a few hundred $\mu$Hz, and then assuming that the differences between the smoothed observed spectrum and the model followed a normal distribution, which allowed a comparison utilising least squares \citep[see e.g.][]{2008ApJ...682.1370K, karoff2008, 2010ApJ...713L.169C}. Though this works in practice, there are no mathematically formal arguments for that the differences between the smoothed observed spectrum and the model should follow a normal distribution. A method for modelling the observed spectra that is statistically valid and that does not require the spectra to be smoothed with a Gaussian running mean is therefore presented below. 

In order to model the solar acoustic background we make use of maximum likelihood estimators \citep[see e.g.][and references herein]{1990ApJ...364..699A}. The purpose is to calculate a logarithmic likelihood function $\ell$ between $N$ independent measurements $x_i$ -- i.e. power density at a given frequency, and a model $f_i$ given by a set of parameters ${\bf \lambda}$. Adopting the notation of \citet{1994A&A...289..649T} and \citet{1998A&AS..132..107A} the logarithmic likelihood function can be calculated as:
\begin{equation}
\ell=-\sum_{i=1}^N{{\rm ln} \ p(x_i,{\bf \lambda})},
\end{equation}
where $p(x_i,{\bf \lambda})$ is the probability density function. The position of the minimum of $\ell$ in ${\bf \lambda}$ space will give us the most likely value of ${\bf \lambda}$ \citep{1998A&AS..132..107A}. The so-called formal error bars can be calculated as the diagonal elements of the inverse of the Hessian matrix $h$:
\begin{equation}
h_{ij}=\frac{\partial^2 \ell}{\partial \lambda_i \partial \lambda_j}
\end{equation}

In order to calculate the probability density function we use the formulation by \citet{2004A&A...428.1039A} where a binned version of the original spectrum is compared to the model. The formulation by \citet{2004A&A...428.1039A} can be used on any model $f_i$ and not just for calculating the significance of peaks in the spectrum. We thus calculate the binned spectrum $\mathcal{S}_i(x,n)$ by summing $x_i$ over $n$ bins:
\begin{equation}
\mathcal{S}_i(x,n)=\frac{1}{n+1}\cdot \sum_{k=i-n/2}^{k=i+n/2}x_k.
\end{equation}
Following \citet{2004A&A...428.1039A} the probability density function is then given as:
\begin{equation}
p\left[\mathcal{S}_i(x,n),f_i({\bf \lambda})\right]=\frac{\mu_i^{\nu_i}}{\Gamma(\nu_i)}\mathcal{S}_i(x,n)^{\nu_i-1}e^{\mathcal{S}_i(x,n)},
\end{equation}
where $\Gamma$ is the Gamma function and $\mu_i$ and $\nu_i$ are given as:
\begin{equation}
\mu_i=\frac{(n+1)\displaystyle\sum\limits_{k=i-n/2}^{k=i+n/2}{f_k({\bf \lambda})}}{\displaystyle\sum\limits_{k=i-n/2}^{k=i+n/2}{f_k^2({\bf \lambda})}},
\end{equation}
and
\begin{equation}
\nu_i=\frac{\left[\displaystyle\sum\limits_{k=i-n/2}^{k=i+n/2}{f_k({\bf \lambda})}\right]^2}{\displaystyle\sum\limits_{k=i-n/2}^{k=i+n/2}{f_k^2({\bf \lambda})}}.
\end{equation}
$n$ appears in $\mu$ and $\nu$ in order to get the right amplitudes. 

\subsection{Testing the facular hypothesis}
Using the formalism above the model in eq.~4, either containing or not containing a facular component, can be compared to an observed spectrum. The model containing a facular component is based on a total of 9 free parameters -- i.e. three from granulation, three from faculae and three from the p-mode oscillations, whereas the model not containing a facular component is based on only 6 free parameters. Having the maximum likelihood of each model from eq. 6 we are then able to calculate the statistical significance of the facular component in the acoustic spectrum by calculating the logarithmic likelihood ratio $\Lambda$:

\begin{equation}
{\rm ln} \ \Lambda = \ell(\lambda_{p+q})-\ell(\lambda_{p}).
\end{equation}
In our case $p$ equals 6 and $q$ equals 3. Following \citet{wilks} \citep[see also][]{1998A&AS..132..107A} we can then compare the value of $-2$ln$\Lambda$ to a $\chi^2$ distribution with $q$ degrees of freedom and in this way calculate the significance of the facular component in the acoustic spectrum. 

\begin{table}
\caption{Acoustic background parameters}
\centering
\begin{tabular}{lcc}
\hline \hline
 & faculae & no faculae \\
 \hline
$\sigma_{\rm gran}$ & 73.0  $\pm$ 1.6 ppm & 85.8  $\pm$ 1.5 ppm\\ 
$\tau_{\rm gran}$ & 214.3  $\pm$ 2.9 sec & 219.0  $\pm$ 4.9 sec\\
$\alpha$ & -3.5 $\pm$ 0.3 & -1.8 $\pm$ 0.1 \\ 
$\sigma_{\rm fac}$ & 61.5  $\pm$ 1.9 ppm  & \\
$\tau_{\rm fac}$ &  65.8  $\pm$ 2.5 sec  & \\
$\alpha$ & -6.2 $\pm$ 0.7 & \\
$H_{\rm osc}$ & 0.7  $\pm$ 0.04 ppm$^2$/$\mu$Hz & 0.52  $\pm$ 0.05ppm$^2$/$\mu$Hz\\
$\nu_{max}$ & 3104  $\pm$ 36 $\mu$Hz & 3084  $\pm$ 44 $\mu$Hz \\
$Width$ & 316  $\pm$ 36 $\mu$Hz & 254  $\pm$ 32 $\mu$Hz \\ 
\hline
\end{tabular}
\label{tab1}
\end{table}

\section{Results}
Two kinds of results are presented: One set of results on the average spectrum of the 119 one months substrings and one set of results on the temporal variability in the spectra of the individual substrings. The analysis is made by binning 100 frequency bins ($n = 100$), which gives a frequency resolution in the binned spectra of around 39 $\mu$Hz. Error bars for all parameters were calculated as the diagonal elements of the inverse of the Hessian matrix (see eq.~7)

\subsection{Average results}
The average spectrum from the 119 one month substrings is modelled with a model containing a facular component and a model not containing a facular component. The results from the modelling are given in Table~1 and the average observed spectra with the models overlaid are shown in Figs.~2~\&~3. As the decay rates were treated as free parameters in both models the decay rates for the granulation component are different in the two models. This leads to differences in the models at high frequency as the obtained decay rate $\alpha$ of $1.8$ $\pm$ 0.1 for the granulation component in the model without a facular component is too small to describe the acoustic background at high frequency. 

\subsection{Comparison to high-resolution observations}
A number of studies have used high-resolution observations of the solar surface to measure the amplitude and time-scale of granulation and faculae, but in general there is little agreement between the different results. The reason for this is probably differences in the way the amplitude and the time-scale are measured, but also differences in the resolution of the observations, differences in the solar activity level at the time of the observations and whether the observations were taken close to the disk centre or close to the limb. The high-resolution observations can therefore only be used as a rough comparison to the seismic results.

If a typical size of granules (or cell area) of 1.2 Mm$^2$ \citep{2004A&A...428.1007D} is assumed, then around 2.5 million granules are found on the visible solar surface. The contrast of a typical granule is around 20\% \citep{2009A&A...503..225W}. The typical contrast of faculae is around 50\% at the peak distance from the disk center ($\mu \simeq 0.6$) \citep{2007ApJ...661.1272B} and the relative fraction of granules that show faculae is around 1.8\% \citep{1995ApJ...454..531B}.  

If  each granule on the solar surface is assumed to cause a signal of relative amplitude $\frac{\delta L}{L}$ at a random time and the entire solar surface is covered with granules, then this will cause a time series of integrated Sun light with a variance of \citep{2006A&A...445..661L, 2009A&A...506..167L}:
\begin{equation}
\sigma \simeq \frac{ \left( \frac{\delta L}{L} \right)_{granule} }{\sqrt{N_{granule}}},
\end{equation}
where $N_{granule}$ is the number of granules present on the visible solar surface at any given time. 

The amplitudes of the granulation component were measured to be $73.0 \pm 1.6$ and $85.8 \pm 1.5$ ppm for the model containing a facular component and the model not containing a facular component, respectively. This is consistent with the theoretical estimate of around 126 ppm that is found by dividing 20\% with the square root of 2.5 million granules. 

The amplitude of the intensity signal from faculae follows that of granules, except that it cannot be assumed that the entire solar surface is covered with faculae. In the most simple approach the amplitude should therefore be scaled with the relative coverage of faculae $\frac{N_{faculae}}{N_{granule}}$:
\begin{equation}
\sigma \simeq \sqrt{\frac{N_{faculae}}{N_{granule}}} \frac{ \left( \frac{\delta L}{L} \right)_{faculae} }{\sqrt{N_{granule}}},
\end{equation}

The amplitude of the facular component was measured to be $61.5  \pm 1.9$ ppm. This agrees roughly with the theoretic estimate of around 42 ppm obtained by multiplying $\sim$50\% with the square root of $\sim$1.8\% and dividing it with the square root of 2.5 million granules.

The characteristic time-scales of convection were measured to be $214.3  \pm 2.9$ and $219.0  \pm 4.9$ sec for the model containing a facular component and the model not containing a facular component, respectively. This is consistent with the coherence time-scale between 156 and 246 sec found by Del Moro (2004). For faculae a characteristic time-scale of $65.8  \pm 2.5$ sec was measured which is consistent with the fact that it should be shorter than the granulation time-scale.

The amplitude and characteristic time-scale of the granulation and facular components in the acoustic spectra are consistent with the high-resolution measurements. This indicates that the interpretation that the second bump in the acoustic spectrum originates from faculae is consistent -- though the comparison does not make the argument alone. The interpretation is also based on the relative relations between the amplitude and characteristic time-scale of granulation, faculae and p-mode oscillations.

Eq.~13 shows why the component in the acoustic spectrum between the granulation and p-mode oscillation components does not originate from a second granulation population as suggested by \citet{2005A&A...443L..11V}. This suggestion is based on the work by \citet{2004A&A...428.1007D} who finds, using high-resolution images of quiet granulation, a second population of granules with a shorter coherence time, but also smaller brightness contrasts. The fact that this second population of granules is both more numerous and have smaller brightness contrasts than ordinary granules means that it should not be visible in the acoustic spectrum as the numerator of eq.~13 would be smaller and the denominator would be larger than for ordinary granulation.

\begin{figure}
\includegraphics[width=\columnwidth]{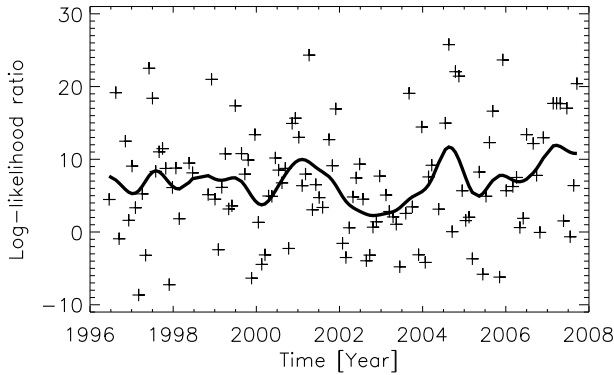}
\caption{The logarithmic likelihood ration between the model with and without the facular component as a function of time. No significant variability following the solar cycle is seen. The solid black line shows a Gaussian running mean with a width ($\sigma$ not FWHM) of 0.4 year.}
\end{figure}
\subsection{Statistical significance of the facular signal}
The significance of the facular component in the average observed spectrum can be calculated from the likelihood ratio between the two different models. The significance was measured to be 83\%. This means that though the model with the facular component and a total of 9 free parameters does describe the average observed spectrum better than the model without the facular component and 6 free parameters, it does not describe it significantly better. Visually, it is clear that the fit in Fig.~2 is better than the fit in Fig.~3, but given the statistic of the observed spectra and the number of free parameters it is not significant better.

In order to analyze why the facular component is not significant in the average observed spectrum we calculate the significance of the facular component in the spectrum of each individual one month substring. The decay rates and characteristic time-scales for the granulation and facular components and the frequency of maximum power for the p-mode oscillation component had to be fixed to the values given in the first column of Table~1. If this was not done the location and slope of the exponential decaying function would change in order to make up for changes in amplitude. This means that the models used for modelling the 119 individual one month substrings contained  3 and 4 free parameters, respectively. 

\begin{figure}
\includegraphics[width=\columnwidth]{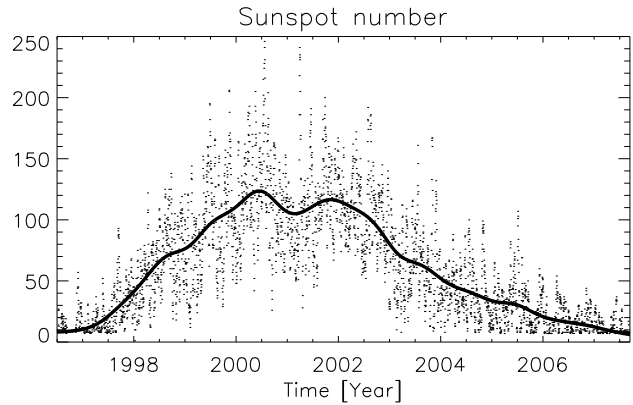}
\caption{Temporal variability of the sunspot number for comparison. The figure shows the international sunspot number obtained from the Solar Influences Data Analysis Center at the Royal Observatory of Belgium. The solid black line shows a Gaussian running mean with a width of 0.4 year, as in the other figures.}
\end{figure}
The logarithmic likelihood ratios between the models with and without a facular component for the 119 individual one month substrings are shown in Fig.~4. It is seen that most of the logarithmic likelihood ratios are between 0 and 10 with a few scatter points below and above. The figure does reveal a few points with logarithmic likelihood ratios below 0. A negative logarithmic likelihood ratio means that in these months the model without the facular component describes the observed spectrum better than the model with the facular component, or in other words, that adding more free parameters to the model makes a worse agreement between the observed spectrum and the model for these months. The reason for this is that the decay rates of the granulation components are not the same in the two different models and for some months the slow decay rate in the model without the facular component describes the observed spectrum better than the fast decay rate in the model with the facular component. These scatter points with logarithmic likelihood ratios below 0 could be the reason for the facular component not being significant in the average observed spectrum.

Clearly, a lot of variability is seen in Fig.~4, but no trend known to follow the visibility or amplitude of faculae is seen. In particular no clear variability that follows the solar cycle is seen (see Fig.~5 for comparison). The reason for this could either be that the measured ratios are not accurate enough to show intrinsic facular variability or that the acoustic signal from faculae depends stronger on the number of available granules than on the fraction or contrast of faculae (see eq. 14).

Fig.~6 shows the distribution of the significances of the facular signal measured in the 119 one month substrings assuming that the logarithmic likelihood ratios follow a $\chi^2$ distribution with one degree of freedom. This distribution shows a clear increase towards a significance of one that falls of rapidly down to a significance of around 0.97 and then have a long tail down to smaller significances (also smaller significances than what is shown on the figure). The conclusion from the figure is that the acoustic signal from faculae is significantly present in the observed solar acoustic background 47\% of the time.

This figure support our impression that the acoustic signal from faculae is present on the Sun most of the time and that the reason that the average spectrum does not shows a significant facular signal is because of a few months where the significance is really low -- i.e. where the model without the facular component describes the observed spectrum better. The same effect is seen in the spectrum calculated from the full 13 years of observations -- i.e. that the acoustic signal from faculae in not significantly present in the spectrum. Again, the reason for this is most likely that the signal is not present all the time, which reduces the significance of the signal when analyzing the full 13 years of observations.

\begin{figure}
\includegraphics[width=\columnwidth]{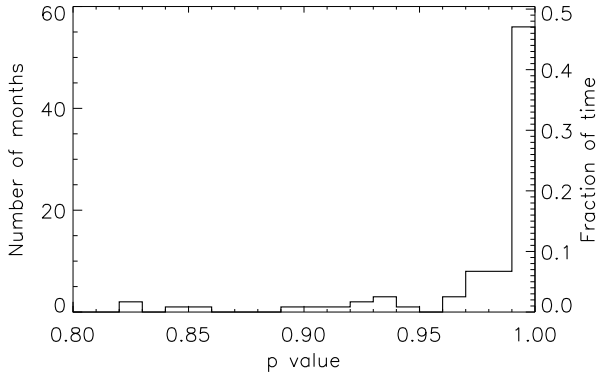}
\caption{Histogram of the measured significances of the facular component in the 119 one months substrings. The significances are calculated using the logarithmic likelihood ratios in Fig.~4 and assuming that these ratios follow a $\chi^2$ distribution with one degree of freedom. It is seen that 56 months (47\%) have a significances higher than 0.99, while 75 (63\%) have a significance higher than 0.95.}
\end{figure}
\subsection{Temporal variation of p-mode oscillations}

\begin{figure}
\includegraphics[width=\columnwidth]{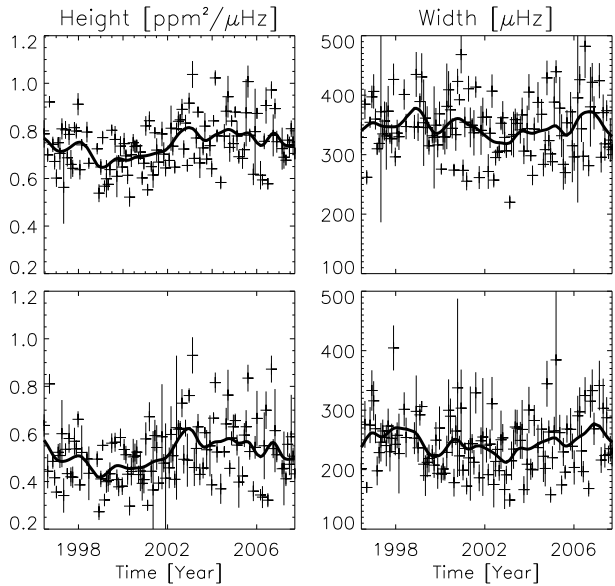}
\caption{Temporal variability of the height (left) and width (right) of the p-mode oscillation envelope for the model with a facular component (top) and the model without a facular component (bottom). The solid black lines show a Gaussian running mean with a width of 0.4 year. As expected a clear anti-correlation is seen between the measured height of the p-mode oscillation envelope and the solar cycle.}
\end{figure}
Both the frequencies, amplitudes and line-widths of the individual p-mode oscillation modes in the acoustic spectrum are known to change with the solar cycle \citep[see e.g.:][]{1990Natur.345..779L, 2000MNRAS.313...32C} and we therefore also expect the average shape and position of the p-mode oscillation envelope to change with the solar cycle. The temporal variability of the height and width of the p-mode oscillations envelope is shown in Fig.~7 for the model with and without a facular component (the position of the envelope is kept fixed in both models). There is a good agreement between the results from the two different models -- with a linear Pearson correlation coefficient of 0.95 (0.97 for the smoothed curves) for the height and 0.83 (0.70 for the smoothed curve) for the width. A clear anti-correlation is seen between the height of the envelope and the solar cycle (see Fig.~5 for comparison) so that the height has minimum around 2000, which coincides with the peak of the solar cycle. The anti-correlation between the height of the envelope and the solar cycles is consistent with other studies \citep[i.e.][]{2008ApJ...682.1370K, 2010Sci...329.1032G} and can be explained by the fact that  the life-time of the p-mode oscillation is known to decrease with increasing solar activity \citep{2003ApJ...582L.115C}. The envelope width does not show any correlation with the solar cycle. The position of the envelope, which is not analyzed here, as it was fixed to the average value, does only show small changes compared to the hight and width \citep{2010Sci...329.1032G}.

The uncertainties on the parameters measured in the average spectrum are comparable in size to the uncertainties measured in the individual one month substrings. This reflects that the scatter seen in i.e. Figs. 7 \& 8 is much larger than the uncertainties on the individual measurements -- in other words it reflects that the parameters does show intrinsic variability on a monthly time-scale.

\subsection{Temporal variation of granulation}
The temporal variability of the amplitude of the granulation component does not show any clear correlation with the solar cycle, neither for the model with or without the facular component. The amplitudes of the granulation component show an offset between the two different models and only some agreement is seen in the temporal variability -- with a linear Pearson correlation coefficient of 0.74 (0.84 for the smoothed curves). Instead some signs of a quasi-annual modulation is seen, especially in the results from the model with the facular component. The ratio between the variance of the measurements (the scatter) and the mean value of the uncertainties is 5.12 for the model with and 4.97 for the model without the faculae component, indicating that the variability is not due to noise. A quasi-annual modulation of both the granulation characteristic time-scale and amplitude was also seen in the analysis of observations from the GOLF instrument on $SOHO$ by \citet{2008A&A...490.1143L} in phase with the variability seen here. \citet{2008A&A...490.1143L} explained the modulation by the fact that the observation altitude (or depth) of the GOLF instrument in the solar photosphere changes with the observation wavelength which again changes with the orbital velocity of $SOHO$, which changes with a period of one year. As the VIRGO observations analyzed here are intensity observations, the effect of a wavelength change should not be prominent, but  the quasi-annual variability is still seen, especially around solar activity maximum. 

\begin{figure}
\includegraphics[width=\columnwidth]{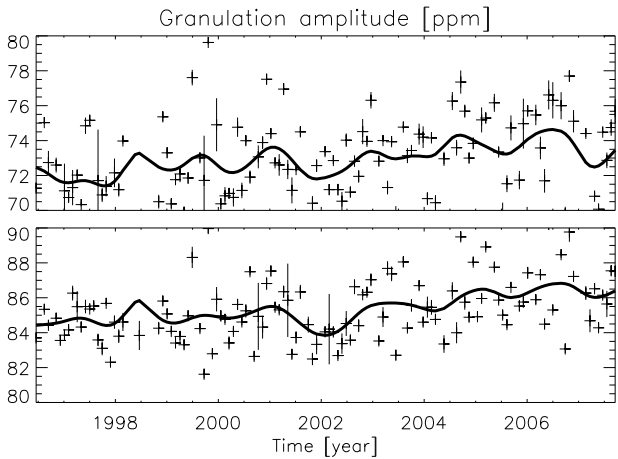}
\caption{Temporal variability of the amplitude of the granulation component for the model with a facular component (top) and without a facular component (bottom). Again, the solid black lines show a Gaussian running mean with a width of 0.4 year. Some signs of a quasi-annual variability is seen, especially in the results from the model with the facular component, as was also the case in the study by \citet{2008A&A...490.1143L}.}
\end{figure}

\subsection{Temporal variation of faculae}
Though the facular component is not significantly present in the acoustic spectra of all the one month substrings the amplitude of the component can still be measured in all the substrings. These amplitudes are plotted in Fig.~9. The uncertainties are relatively large for these measurements, and the measured amplitudes show only little scatter within these uncertainties. The amplitude of the scatter is around a few ppm. The Gaussian running mean of the measurements (represented with the black line) show a quasi-biennial variability with maximum around 2001, 2003, 2005 and 2007. The amplitude of this variability is around one ppm -- i.e. significantly lower than the scatter of the individual points. It is therefore not clear if the quasi-biennial variability is intrinsic to the Sun or if it is just random noise. 

A similar quasi-biennial variability was seen in the residuals of the p-mode oscillations frequency shifts in the study by \citet{2010ApJ...718L..19F}. In Fig.~10 we therefore compare the residuals of the measured amplitudes of the facular component with the residuals of the p-mode oscillations frequency shifts from \citet{2010ApJ...718L..19F}. In order to do that we have subtracted a 3rd order polynomial from the measurements in Fig.~9 and binned the one month measurements to the temporal resolution of the residuals of the p-mode oscillations frequency shifts (182.5 days). The two sets of measurements show a correlation with a linear Pearson correlation coefficient of 0.34. The reason that no strong correlation is seen could however be the relative large uncertainties on both sets of measurements and it can thus not been ruled out that the faculae and the p-mode oscillations frequency shifts have the same origin.

\citet{2010ApJ...718L..19F} speculate that the quasi-biennial variability seen in the residuals of the p-mode oscillations frequency shifts originates from a second solar dynamo located not at the so-called tachocline at the base of the convection zone, but in the strong shear layer just below the surface. This idea of a near surface dynamo, which has also been suggested by \citet{2005ApJ...625..539B}, could also explain the short periodic magnetic variability that has recently observed in other Sun-like stars \citep{2007ApJ...657..486B, 2010Sci...329.1032G, 2010ApJ...723L.213M}. If the presents of faculae and p-mode oscillation frequency shifts are indeed correlated (which we can however not claimed based on the present measurements) then this could suggests that the presence of faculae on the surface of the Sun is controlled not only by the 11-years solar cycles, which is expected to be located in the solar tachocline, but also by a quasi-biennial dynamo located in the strong radial shear layer just below the solar surface \citep{2005ApJ...625..539B}

\begin{figure}
\includegraphics[width=\columnwidth]{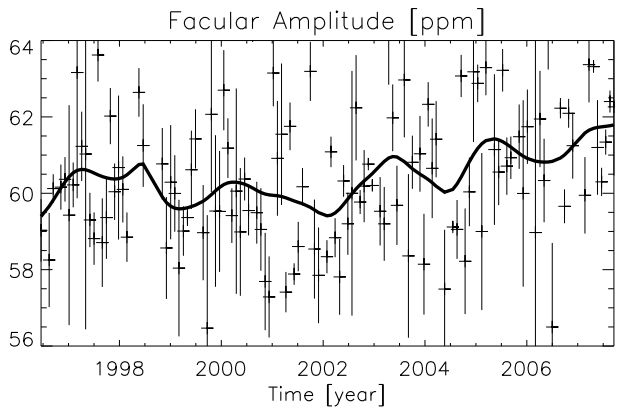}
\caption{The temporal variability of the amplitude of the facular component. Again, the solid black lines show a Gaussian running mean with a width of 0.4 year. Though the measured amplitudes show only little scatter within the uncertainties the black line reveals a quasi-biennial variability. A similar quasi-biennial variability is also seen in the residuals of the p-mode oscillations frequency shifts in the study by \citet{2010ApJ...718L..19F}}, which suggest that the variability in the amplitude of the facular component could be caused by a second dynamo.
\end{figure}

In general it has to be noted that the quasi-annual and quasi-biennial periods, seen in the granulation and facular amplitudes, are very peculiar. Varying instrumental effects such as altitude, pointing, temperature or the like could be the culprit. One of the things that were done in order to test this was to redo the analysis using only the monthly substrings where the facular signal had a significance larger than 95 and 99\% and this did not change any of the conclusions on the facular signal - i.e. the same kind of periodicity were still seen.

\section{Conclusion}
13 years of observations of the integrated brightness of the Sun from the VIRGO instrument on $SOHO$ have been analyzed. Three distinct features are seen in the frequency spectrum of these observations. By comparing the measured characteristic time-scale and amplitude of these phenomena to high-resolution observations of the solar surface it is shown that the three features can convincingly be explained as granulation, faculae and p-mode oscillations.

The feature in the acoustic background related to faculae is the weakest and it is not significantly present in the average spectrum made from 119 individual one month spectra. The feature is on the other hand present in 56 (47\%) out of the 119 one month spectra with a significance larger than 0.99 and in 75 (63\%) with a significance larger than 0.95. 

The temporal variability of the amplitudes of the granulation, facular and p-mode oscillations components in the solar acoustic background have also been analyzed. The p-mode oscillations show a clear anti-correlation with the solar cycle and the amplitude of the granulation component show some signs of an quasi-annual variability, which is also seen in velocity observations \citep{2008A&A...490.1143L}, but no correlation with the solar cycle. The amplitude of the facular component does also not show any correlation with the solar cycle, but instead some signs of a quasi-biennial variability is seen. The residuals of the temporal variability of the amplitude of the facular component show a correlation with a linear Pearson correlation coefficient of 0.34 with the residuals of the p-mode oscillations frequency shifts from the study by \citet{2010ApJ...718L..19F}.

\begin{figure}
\includegraphics[width=\columnwidth]{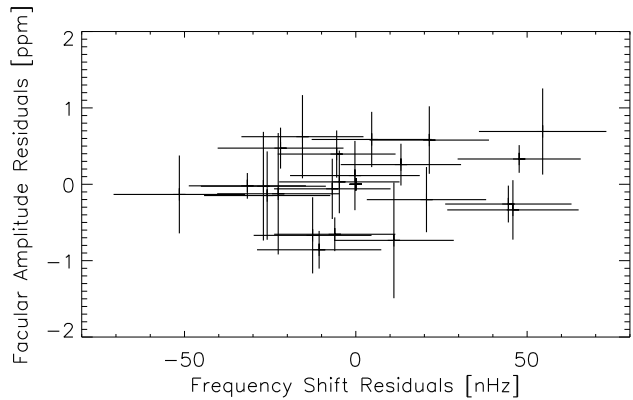}
\caption{Relation between the residuals of the measured amplitude of the facular component and the residuals of the p-mode oscillations frequency shifts from the study by \citet{2010ApJ...718L..19F}. The two sets of measurements have a linear Pearson correlation coefficient of 0.34, which means that no strong correlation can be claimed between the amplitude of the facular component and the p-mode oscillation frequency shifts. The reason for this could be the relative large uncertainties on both sets of measurements and it can thus not been ruled out that the faculae and the p-mode oscillations frequency shifts have the same origin.}
\end{figure}
\section*{Acknowledgments}
I would like to thank W.J.~Chaplin and H.~Kjeldsen for useful discussions, J.~Christensen-Dalsgaard for suggesting that I wrote this paper and T.~Appourchaux for patiently explaining to me how the significance of the facular signal could be tested. I also thank A.B. Broomhall for providing me with the residuals of the p-mode oscillations frequency shifts from BiSON. The Carlsberg foundation is acknowledged for financial support.

\appendix

\section[]{Monte Carlo simulations of the acoustic signal from faculae}
In order to test the robustness of the statical analysis developed in Section 3, a number of Monte Carlo simulations were made. In the simulations it was assumed that the acoustic spectrum followed the model given in eq.~4 and an acoustic spectrum was simulated for the parameters given in Table~1. Using an inverse Fourier transform the acoustic spectrum was convented from the frequency to the time domain, where realistic random noise were added at each time-step. Hereafter the acoustic spectrum was convented back into the frequency domain, where it was analyzed -- the background parameters and the significance of the facular signal measured.

The two parameters that were changed in the simulation were the amplitude of the facular component and the noise level -- all other parameters were fixed to the values given in Table~1. This means that for each set of facular amplitude and noise level the simulation returned not only measured facular amplitude and noise level, but also measured values for all the other parameters in Table~1, which could be used to evaluate the precision of the procedure. The amplitude of the facular component was changed between 45 and 70 ppm and the noise level from 0 to 80 ppm per measurement (equivalent to the point-to-point scatter in the time series).

\begin{figure}
\includegraphics[width=\columnwidth]{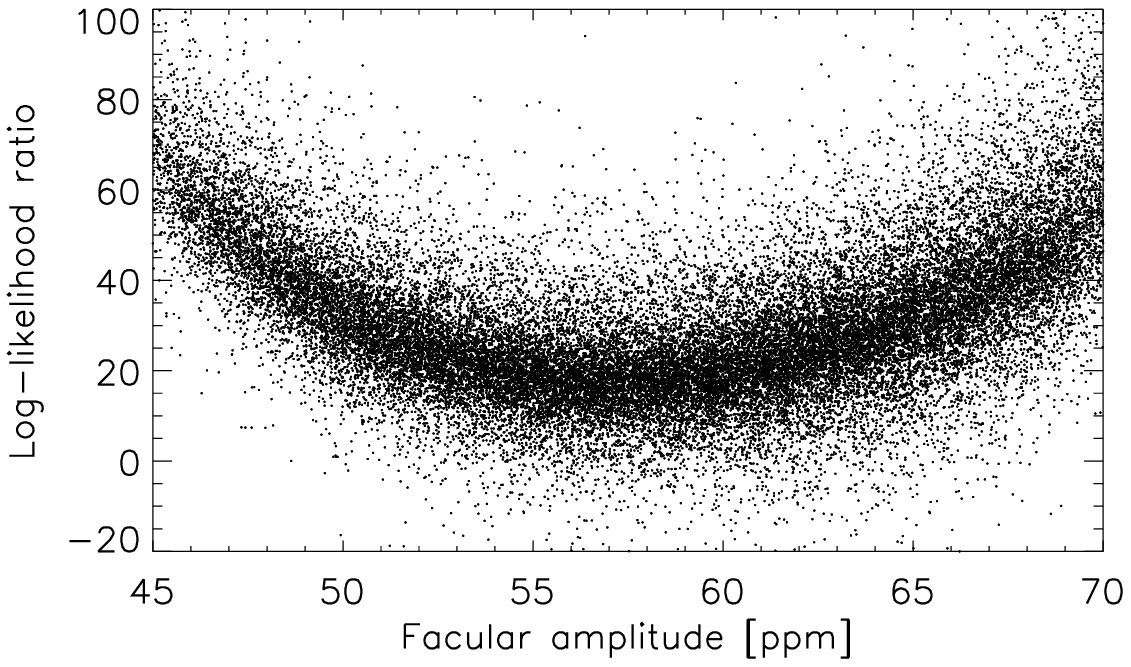}
\caption{Relation between the measured amplitude and the measured significance of the facular component in the simulations (here just given as the logarithmic likelihood ratio in order to enhance the structures in the plot). It is seen that the relation is far from linear. The reason for this is, that the likelihood between the simulations and the model that does not include a facular component is not a linear function of the amplitude of the facular component. The likelihood between the simulations and the model that does include a facular component is on the other hand linear.}
\end{figure}
The main result of this Monte Carlo simulations is shown in Fig.~A1, which shows the relation between the amplitude of the facular component and the logarithmic likelihood ration between the model with and without the facular component. It is seen that there is no linear correlation between the two. This is a bit surprising as a spectrum with a hight amplitude facular component is expected to have a higher likelihood with a model with a facular component than a spectrum with a low amplitude facular component would have. This is also the case, but the problem is that there is no linear relation between the likelihood and the amplitude of the facular component for the model that does not include the facular component. In other words, is the reason why no linear correlation is seen that though the likelihood between the model and the observations increases linear as a function of facular amplitude for the model with the facular component, the same (or more correctly the opposite) is not true for the model without the facular component. 

This does first of all explain why some sings of variability can be seen in Fig.~9, which shows the measured facular amplitude as a function of time, while not in Fig.~4, which shows the likelihood ration as a function of time.

It does also explain why we can measure the amplitude and characteristic time-scale of the facular component and the uncertainties on these parameters even in months where the component is not significant. The reason is that in order to measure the amplitude and characteristic time-scale we only use the model that includes the facular component, while in order to measure the likelihood ratio we also uses the model that does not include the facular component.

No trends were seen between the measured parameters and the noise level in the simulations, except for increased scatter. For some of the simulations with the highest noise levels it was not possible to get the model to convert to the observed spectrum. 

\begin{figure}
\includegraphics[width=\columnwidth]{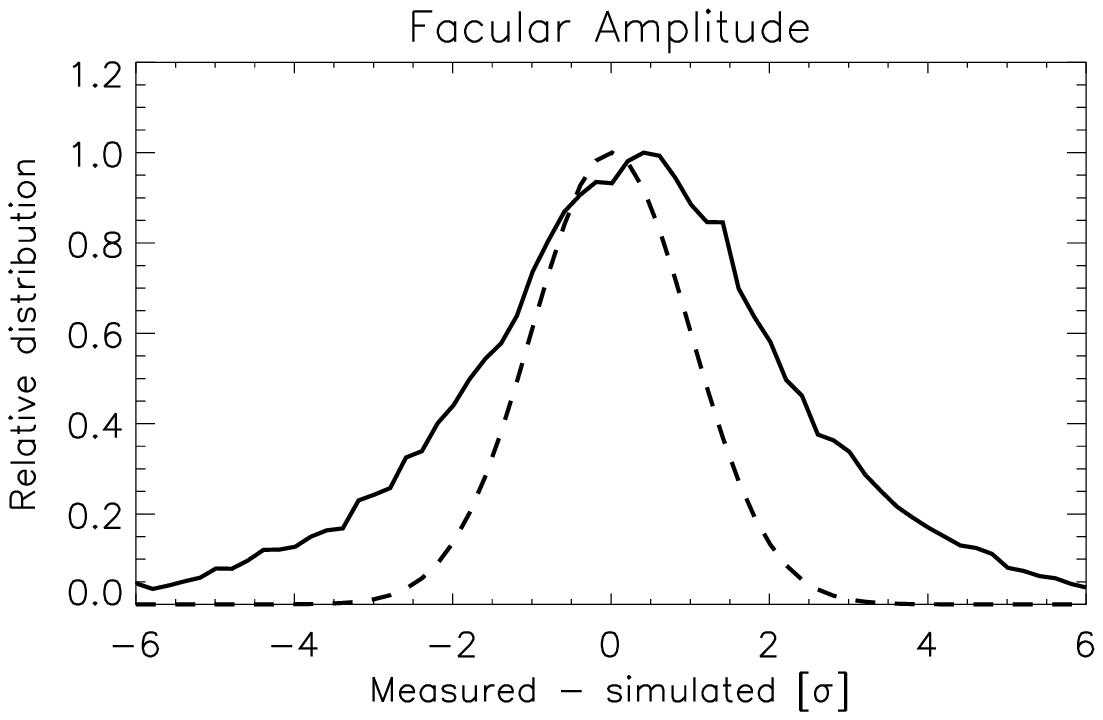}
\caption{Distribution of the difference between the input and output of the simulations. The figure shows a histogram of the difference between the facular amplitude that was used in the simulations and the amplitudes that were returned from the model, in therms of the uncertainty returned from the modeling of the simulated spectra (solid line). The dashed line shows a Gaussian with a variance of one for comparison. The fact that the distribution is not to different from a Gaussian with a variance of one, reflects that the uncertainties returned from the modelling are realistic.} 
\end{figure}
The Monte Carlo simulations have also been used to test the uncertainties on the different parameters - especially the parameters related to the facular component. This is done by plotting a histogram of the difference between the parameter values used in the simulation and the measured parameter values divided by the uncertainty on the measured parameter values ($\frac{\lambda_{obs}-\lambda_{true}}{\sigma_{\lambda}}$). If the errors on the parameters were random distributed around a mean value we would expect that these histograms would all be Gaussian function with a variance of one. This is also the case for the parameters in Table~1 (the parameters related to the facular component, is shown in Fig.~A2), though the variance might be a bit larger than one, this is not unexpected as the uncertainties are only formal uncertainties. In order to get more reliable uncertainties one would eventual have to preform a full Bayesian analysis as it is sometimes done when modeling the p-modes using Markov chain Monte Carlo \citep[see e.g.][]{2011A&A...527A..56H}.

\label{lastpage}

\end{document}